# Anomalous spectral shift of near- and far-field plasmonic resonances in nano-gaps

*Anna Lombardi[1], Angela Demetriadou[2,3], Lee Weller[1], Patrick Andrae[1†], Felix Benz[1], Rohit Chikkaraddy[1], Javier Aizpurua[2], Jeremy J. Baumberg*[1]*

[1] NanoPhotonics Centre, Cavendish Laboratory, University of Cambridge, Cambridge, CB3 0HE, UK
[2] Centro de Física de Materiales, Centro Mixto CSIC-UPV/EHU and Donostia International Physics Center (DIPC), Paseo Manuel Lardizabal 4, 20018 Donostia-San Sebastián, Spain
[3] Blackett Laboratory, Department of Physics, Imperial College London, SW7 2AZ, United Kingdom

## Abstract

**The near-field and far-field spectral response of plasmonic systems are often assumed to be identical, due to the lack of methods that can directly compare and correlate both responses under similar environmental conditions. We develop a widely-tuneable optical technique to probe the near-field resonances within individual plasmonic nanostructures that can be directly compared to the corresponding far-field response. In tightly-coupled nanoparticle-on-mirror constructs with nanometer-sized gaps we find >40meV blueshifts of the near-field compared to the dark-field scattering peak, which agrees with full electromagnetic simulations. Using a transformation optics approach, we show such shifts arise from the different spectral interference between different gap modes in the near- and far-field. The control and tuning of near-field and far-field responses demonstrated here is of paramount importance in the design of optical nanostructures for field-enhanced spectroscopy, as well as to control near-field activity monitored through the far-field of nano-optical devices.**

**TOC Graphic**

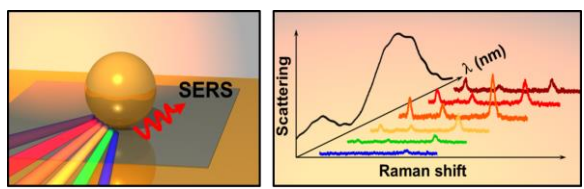

---

[†] Current affiliation: Nanooptical Concepts for PV, Helmholtz-Zentrum Berlin, Hahn-Meitner-Platz 1, 14109 Berlin, Germany

The interaction of light with noble metal nanostructures excites collective electron oscillations in the form of localized plasmonic resonances. As a result, such plasmonic nanostructures are able to confine light within extremely small volumes, millions of times smaller than a wavelength-sized box. Squeezing light into such small regions creates near-thousand-fold field enhancements, which are ideal for intense surface-enhanced Raman scattering (SERS), thus allowing only a few atoms, molecules or nano-objects to be directly tracked.(1) So far, researchers have typically assumed that both the localised near-field and radiated far-field support their resonant behaviour (i.e. strongest field enhancements) at similar spectral wavelengths. As a result, optimisation of SERS has depended on measurements of the far-field scattering spectrum.

Here we show that when the optical field is tightly confined by nanoscale gaps the resulting multiple order plasmon resonances supported at different wavelengths interfere with each other differently to build up the signal from the near- and far-fields. As a result, significant spectral shifts are observed. We experimentally demonstrate this using a spectral-scanning technique that simultaneously records dark field scattering spectra and tuneable-pump SERS measurements on each nanostructure individually. We utilise plasmonic constructs for this which provide extremely robust nanoscale gaps,(2) using Au nanoparticles separated from a bulk Au film by an ultrathin molecular spacer, known as the nanoparticle-on-mirror (NPoM) geometry.(3,4,5) In contrast to the red-shifts always found in isolated nanoparticles,(6) the near-field NPoM resonance from SERS is found to be always blue-shifted from the scattering peak. We explain this through a transformation optics model which allows the decomposition of the total signal into individual modes which show different radiative properties. In particular the $n$=1 and 2 modes interfere constructively in the far-field, but destructively in the near-field. From this understanding, our experiments also allow us to show that the SERS background arises from a completely different process to SERS, as it follows instead the far-field spectral enhancement. Our insights provide a solid intuition to predict how near-fields behave within a wide variety of plasmonic nano-constructs.

Direct measurements of the near-field plasmonic enhancement spectra are either probe-based techniques, or must exploit nonlinear processes, since only then do the evanescent fields contribute most strongly. Probe-based techniques are not suitable for single NPoM measurements and second harmonic generation is not very reliable for this task as it possesses both bulk and surface contributions and is thus very sensitive to many additional aspects of nanoscale geometry. Third-harmonic generation techniques(7) are also possible but so far are primarily single wavelength studies.(8,9,10,11,12) The other favourable process for this task is SERS, but this has also been challenging because of the requirement for wide tuning of the Raman pump laser, while ensuring high-contrast tuneable filtering of the scattered light from the background Rayleigh pump scatter. As a result most experiments work with arrays of nanoparticles(13) or use a limited number of excitation wavelengths on individual nanostructures.(14,15,16,17,18) Alternative approaches with fixed excitation wavelength which attempt to tune the plasmon resonance suffer uncontrolled changes in confinement and enhancement.(8,9,10,11,12) Recent experiments(15) have managed to deliver wavelength-scanned Raman and dark-field measurements on lithographically-defined plasmonic dimers in order to ascertain how quantum tunnelling affects the SERS amplitude. Lithography however generates considerable uncertainty in the gap sizes and morphologies. Compared to such nanoparticle dimers, the NPoM geometry guarantees much better control of gap size between gold film and nanoparticle, higher reproducibility, and a much simpler and robust nano-assembly procedure, and has thus been recently utilised in many experimental studies.(3,4,5) The well-defined architecture also precisely defines the orientation of the optical fields and of the molecules which are currently studied in SERS, and thus allows precise comparison of the near- and far-field response.

Our experimental setup is optimized to realize *both* dark field microscopy and broadband tuneable SERS measurements on the *same* single nanoparticle at the same time (Fig.1a). For dark field scattering spectroscopy white light is focused on the sample through a high numerical aperture (NA=0.8) 100× objective and a cooled spectrometer detects the scattering of single nanoparticles which are kept well spatially separated (coverage < 1 µm$^{-2}$).

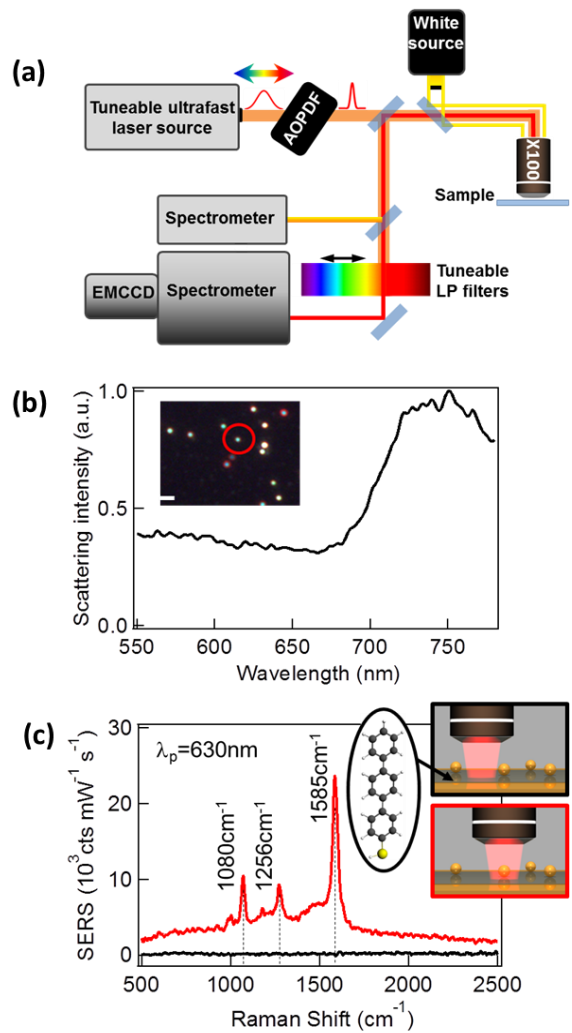

Figure 1: (a) Broadband tuneable SERS spectroscopy coupled with dark field microscopy for single nanoparticle studies, based on AOPDF spectral filter. (b) Scattering spectra from a single 60nm gold nanoparticle on mirror covered with TPT. Inset: dark field image of sample highlighting nanoparticle under study (red circle), scale bar is 1µm. (c) Surface enhanced Raman spectra measured on (red), and off (black), single nanoparticle. Laser

**excitation wavelength is 630nm. Inset: gold nanoparticles on gold substrate, separated by 1.4nm TPT molecular spacer.**

To realise broadband tuneable SERS a sub-nm linewidth tuneable laser source is required. To create this a 200fs Ti:sapphire oscillator pumps a fs optical parametric oscillator to give tuneable output over visible and near-infrared wavelengths from 500nm to 1040nm (see Methods). The output is spectrally narrowed below 1nm using an acousto-optic programmable dispersive broadband filter (AOPDF),(19) yielding fully automated tuning with multi-mW output powers. This Gaussian beam is focussed in an inverted microscope to a diffraction limited spot using the same 100× objective. For each excitation wavelength, Rayleigh scattered light is filtered out using a computer-controlled translatable custom-built array of multiple linear variable long wave pass (LP) filters with an overall optical density OD 11, maximum transmission of 80%, and cut-off spectral width of 10nm. Stokes Raman signals are recorded with a spectrometer and cooled EMCCD camera. Calibration on bulk solids confirms this system is capable of Raman measurements across the entire visible and infrared.

For the near-field measurements here, a *p*-terphenylthiol (TPT) self-assembled molecular monolayer (SAM) is used as a nano-scale spacer between the flat gold surface and 60nm gold nanoparticles placed on top. The gap thickness, which depends on the orientation angle of the molecules on the gold surface, is found to be $d$=1.4±0.1nm through phase modulated ellipsometry measurements, in good agreement with previous work.(1) Individual nanoparticles are first optically characterized by dark field spectroscopy (Fig.1b) which shows for all spectra a strong coupled plasmon mode resonance in the near-infrared around 730±20nm. The <20nm (FWHM) variation in spectral peak arises from the 10% distribution of Au nanoparticle diameters. The tightly confined hotspot created within the gap (lateral size $\sqrt{Rd} \sim$ 6.5nm) is then an ideal situation to compare near- and far-field spectra.

When the laser (at pump wavelength $\lambda_p$=630nm in Fig.1) is focussed away from any Au nanoparticle, the Raman scattering from the molecular SAM is below our detection noise level (Fig.1c, black). However focussing on a NPoM elicits hundreds of times stronger optical fields greatly enhancing the SERS signal of TPT molecules confined within the plasmonic nanogap (Fig.1c, red). Three dominant vibrational lines are seen corresponding to a C-H rocking mode (1080cm$^{-1}$) and to in-plane stretching of the benzene rings (1256cm$^{-1}$ and 1585cm$^{-1}$). The average laser power at the sample is kept below 1µW which is needed to avoid any shifts in the NPoM dark field spectra or any changes in SERS over time (Supp.Info.). Similar SERS signals are obtained from each NPoM.

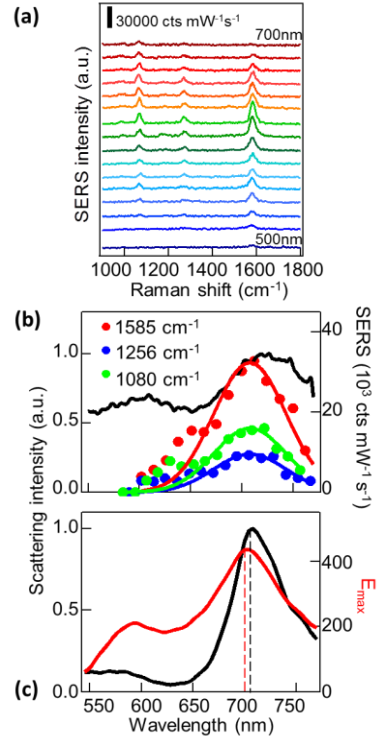

**Figure 2:** (a) SERS spectra from a single 60nm nanoparticle tuning the excitation laser from $\lambda_p$=550-700nm in 10nm steps. (b) Extracted SERS intensity of three TPT Raman peaks *vs* SERS peak (outgoing) wavelength, compared to the normalized scattering spectrum (black) of the same single gold NPoM (lines show Gaussian fits). (c) FTDT simulations of scattering (black) and maximum near-field (red) for a 60nm gold NPoM placed on a gold mirror.

By scanning the laser wavelength, we measure the plasmonic-induced SERS enhancement from TPT to access the near-field spectrum of the NPoM (Fig.2a). The vibrational peaks do not shift with pump $\lambda_p$, but their amplitudes show strong enhancements at $\lambda_p$=640nm (1585 cm$^{-1}$ mode), $\lambda_p$=650nm (1256 cm$^{-1}$ mode) and $\lambda_p$=670nm (1080 cm$^{-1}$ mode) (Fig.2a). Control measurements show the plasmonic origin of these resonances, and imply SERS enhancement factors ~10$^8$ with $N_{SERS}$=200 molecules confined within each hotspot (see Methods).

Plotting the extracted experimental SERS enhancements $I_{SERS}(\nu,\lambda_p)$ of each of the three main TPT peaks against the *outgoing* wavelength (Fig.2b) shows they reach their maxima close to the coupled mode observed in far-field scattering, but are blue-shifted by ~22meV. This is contrary to the behaviour for isolated plasmonic nanoparticles in which case the near-field resonance is found to be red-shifted compared to the scattering.(6,20,21,22) This red-shift is associated with the damping of a plasmonic resonance.(6,22) A localized surface plasmon can be interpreted in terms of a driven damped oscillator. When damping is present the maximum oscillation amplitude occurs at a lower energy than the natural frequency of the oscillator, while maximum dissipation occurs at the natural frequency, giving a spectral red-shift. A more complete description of the oscillator model describing plasmonic resonances was discussed by Kats et al.(23) Blue-shifted near-field spectra were previously reported(24) for large nanoparticle arrays. Using single nanostructures here confirms this behaviour does not originate from any sample inhomogeneity or periodic effect. Finite difference time domain (FDTD) simulations of the maximum near-field and the scattering far-field optical response of a single NPoM confirm this behaviour (Fig.2c) and show it is a general property of all closely-coupled plasmonic resonators such as dimers (Supp.Info.). We note that the overall spectral shifts between simulation and experiment here probably arise from faceting of the nanoparticle, which increases the coupling and red-shifts the coupled plasmon resonance.(25,26)

Results on a range of NPoMs show that the near-field resonance is always blue-shifted from the far-field resonance by 4meV to 55meV depending on the nanoparticle (Fig.3a). Different sized nanoparticles tune the coupled mode, but in all cases $I_{SERS}$ follows a Gaussian spectral profile with a ~two-fold reduction in linewidth compared to the corresponding resonant plasmonic mode in scattering. This is also unexpected since both dark field- and Raman-scattering require each photon to couple in and to couple back out, both resonantly enhanced by the plasmonic antenna.

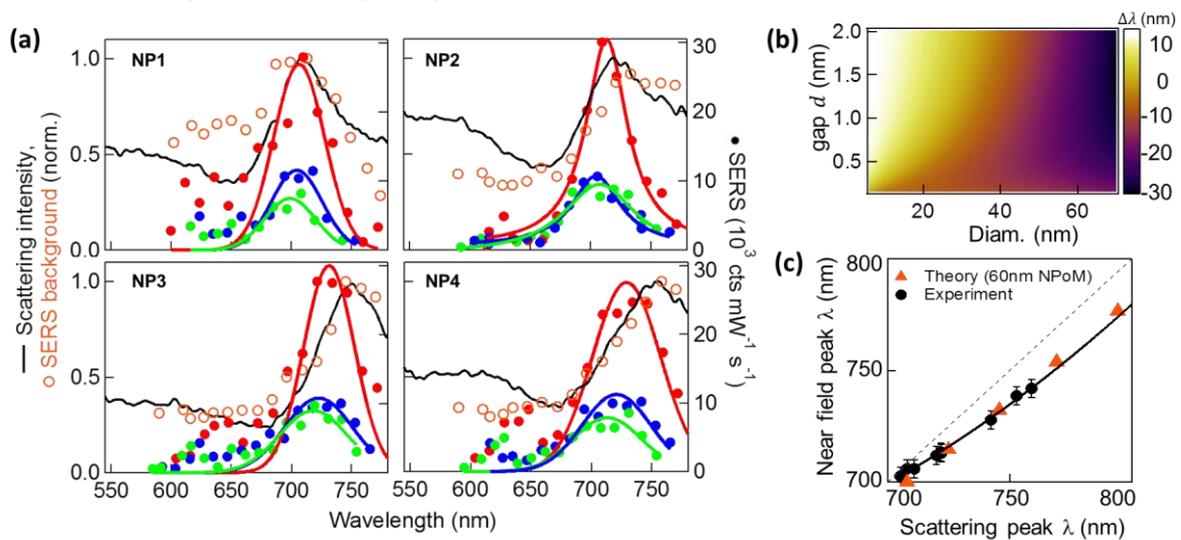

Figure 3: (a) Evolution of SERS intensity of TPT vibrational modes at 1585cm$^{-1}$ (red), 1256cm$^{-1}$ (green), and 1080cm$^{-1}$ (blue) *vs* emitted Raman wavelength, compared to scattering spectra (black) for four NPoMs (lines show Gaussian fits). Also shown is SERS background around 1585 cm$^{-1}$ (orange). (b) Calculated resonance shift $\Delta\lambda = \lambda_{near\ field} - \lambda_{scat}$ for different size nanoparticles in NPoM, and different gap sizes. (c) Comparison of wavelengths of peak SERS emission and the dark field plasmon resonance, for experiments (black) and simulations (orange). Polynomial fit of experimental points is shown as a plain black line.

To understand the origin of this effect and the spectral narrowing of the near-field resonance, it is useful to separate out the contributions from each plasmonic resonance supported in the gap, which thus requires further theoretical insights.

Contrary to an isolated nanoparticle, when two plasmonic resonators (here a nanoparticle and mirror) are coupled, the near-field enhancement is heavily dependent on the physical geometry of the nano-gap and on the excited modes within it. By implementing a transformation optics technique(27) (Supp.Info.), we model the optical response of the gap-modes, both in the near-field and in their radiative emission. The transformation optics technique is developed here in a two-dimensional system to provide the key intuition to understand the nature and composition of the fundamental modes. We find that the dipole localized surface plasmon polariton resonance ($n$=1) strongly interferes with the quadrupole mode ($n$=2) (corresponding field distributions in Fig.S6 of Supp.Info.). This leads to similar field patterns around most of the nanoparticle, but pronounced differences inside the gap (Fig.4a-d). Critically, they have opposite phase, $\phi_n$, in the near-field (i.e. destructive superposition, Fig.4e arrow), but they radiate coherently to the far-field (i.e constructive interference, Fig.4f). Hence, the second gap mode shifts the far-field ($\sigma_{scat}$) resonance to longer wavelengths (Fig.4d) and the near-field (SERS) to shorter wavelengths (Fig.4c).

Even higher-order modes become more significant for extremely small separations (<0.3nm for the NPoM discussed here), where the nanoparticle couples even stronger with the mirror. Their phase shifts alternate with even/odd $n$ in the near-field, but all modes add coherently in the far-field, shifting $\sigma_{scat}$ to even longer wavelengths. Both experiment and simulations confirm this, showing an increased blue-shift for coupled modes which are located further in the infrared. As the nanoparticle moves away from the mirror, the blue-shift observed in the coupled regime decreases until we reach the decoupled regime (isolated nanoparticle) and a red-shift is observed instead (Fig.3b), in agreement with previous results. On the other hand, it has been reported(28,29) that for geometrically more complex structures (such as trimers), the field enhancement in the near field shows a maximum "*in regions where there is no hint of a resonance in the absorption/extinction*".(29) Such structures support multiple modes at nearby frequencies, which commonly result in spectral shifts in the far-field response. Our NPoM system provides a unique opportunity to isolate a precise modal structure, and perform a well-defined modal analysis on a robust spectral composition, which is not typical in other SERS systems. It should be noted that all theoretical calculations are performed purely classically, ignoring non-locality and electron spill-out from the plasmonic metals, the basic concepts here will be little altered by quantum effects for this range of particle-surface separation distances.

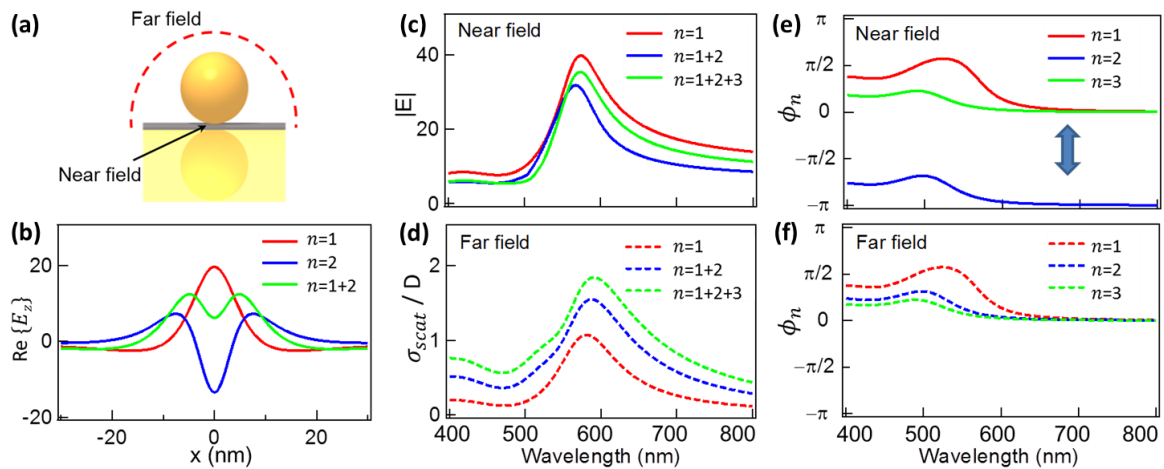

**Figure 4:** (a) Simulation geometry, giving (b) distribution of fields perpendicular to NP surface for dipolar and quadrupolar modes labeled as $n$=1,2. (c,d) Spectral dependence of maximum field in the gap (near-field) and scattering (far-field) for first few $n$. (e,f) Phase of $n$=1,2,3 in the far-field (f) and in the centre of the gap (near-field, e).

Further insights can be extracted from the dependence of the SERS background on $\lambda_p$. This background contribution has been highly debated in the literature,(30,31) with competing explanations of plasmon luminescence, image molecules, inelastic electron scattering, and from contamination, however clarifying data is still lacking. For each NPoM we extract the background in the vicinity of the 1585cm$^{-1}$ peak and plot it as a function of the emitted wavelength (Fig.3a open circles). This clearly shows the SERS background does not match the spectral shape of the near-field enhancement in the gap but closely follows the far-field optical scattering response of the plasmonic NPoM, in both spectral position and linewidth. Recent work(30,31) shows that much of the SERS background must come from optical penetration inside the metal where it can induce inelastic scattering of the electrons. In the experiments here, molecules are placed only in the gap (hence they only probe the near-fields) while the $n$=1,2 modes localise the light around the entire nanoparticle surface. Hence, our spectral tuned measurements thus prove that the SERS peaks and background must arise from different sources, as also suggested by super-resolution imaging studies.(32) As a result we prove that the SERS background observed here has a component that tracks the far-field enhancement, as well as an equally intense spectrally constant component arising from the surrounding planar substrate.

We have thus shown that the clear identification and spectral separation of the near- and far-field resonances can be achieved using precision spectrally-tuned SERS measurements on single nanoparticles. Both experiment and theory agree in the resonance shifts and spectral widths which are found to be controlled by the coherent superposition between different plasmon gap modes. In the spherical nanoparticle-on-mirror geometry the dominant modes are the $n$=1 and 2 which have opposite phase in the near-field but the same phase in far field, resulting in a blue-shift of the SERS peak compared to the dark field scattering, and a two-fold smaller resonance linewidth. This intuitive understanding of how the resonance positions are determined is generally applicable to coupled plasmonic systems. It also shows that the ever-present SERS background does not come from the same spatial locations as the near-field-controlled SERS peaks.

## METHODS

**Sample preparation**
Gold substrates are prepared by evaporating 100nm gold (Kurt J. Lesker Company, PVD 200) on a silicon (100) wafer (Si-Mat, Germany) with a rate of 1Å/s. To obtain atomically smooth films a standard template stripping method is used: silicon substrates are glued onto the freshly evaporated gold using an epoxy glue (EpoTek 377)(33) and the resulting gold / epoxy / silicon sandwich is peeled off the silicon wafer.
Self-assembled monolayers of 1,1′,4′,1″-Terphenyl-4-thiol (Sigma-Aldrich, 97%) are formed by submerging the freshly template stripped substrates into a 1mM solution in water-free ethanol (Sigma-Aldrich, reagent grade, anhydrous) for 24h. The samples are subsequently thoroughly rinsed with ethanol and blown dry. Gold nanoparticles (BBI solutions, UK) are deposited by drop casting from the as-received solution. The deposition time is adjusted in order to obtain the desired nanoparticle coverage. The samples are rinsed with milliQ water in order to remove any salt residues.

**Ellipsometry**

The thickness of the self-assembled monolayers is both measured using ellipsometry (Jobin-Yvon UVISEL spectroscopic ellipsometer) and normalising plasmon resonance spectroscopy.(34) For the ellipsometry measurements an angle of incidence of 70° is used. The data are modelled and fitted using a three layer model. A thickness of 1.5nm is determined with a refractive index of n = 1.45.

**Dark field spectroscopy**

Optical dark field images are recorded on a custom Olympus GX51 inverted microscope. Samples are illuminated with a focused white light source (halogen lamp). The scattered light is collected through a 100× dark field objective (LMPLFLN-BD, NA=0.8) and analysed with a fiber-coupled (50μm optical fiber) Ocean Optics QE65000 cooled spectrometer. We use a standard diffuser as a reference to normalize white light scattering. For each sample, we record optical spectra from 20 randomly selected isolated nanoparticles.

**Tuneable SERS**

An ultrafast laser system based on a 200fs Ti:sapphire oscillator (Spectra Physics MaiTai delivering 200fs pulses, FWHM 10nm, at 80MHz repetition rate) pumps a fs optical parametric oscillator (Spectra Physics Inspire). This light source provides a tuneable output over a wide range of visible and near-infrared wavelengths from 500nm to 1040nm. The monochromaticity of the output beam is reduced below 1nm spectral bandwidth using an acousto-optic programmable dispersive broadband filter (AOPDF, Dazzler, Fastlite). Relying on interactions between a polychromatic acoustic wave and a polychromatic optical wave in the bulk of a birefringent crystal, it is fully automated across a wide wavelength range (500nm-900nm) yielding average output powers of several mW.

SERS experiments are performed on the same modified Olympus GX51 inverted microscope used for dark field spectroscopy. A monochromatic wavelength-tuneable laser beam is focused on the sample using a 100× objective (NA=0.8). Raman scattering is collected through the centre of the objective and analysed with a Shamrock SR-303i fully automated spectrometer coupled with an EMCCD camera water cooled to -85°. For the current experiments we use a 600l/mm 650nm blazed grating. Rayleigh scattering is filtered out with a set of three long pass linear variable filters (DELTA); this filtering system allows the detection of a minimum Raman shift of about 400cm$^{-1}$ over the studied spectral range. The system is calibrated using a silicon substrate as a reference. Spectral acquisitions are taken using an integration time of 10s and an average laser power on the sample below 1μW.

The enhancement factor per molecule (EF) is calculated for each nanoparticle by integrating the Raman peak areas and taking the ratio between SERS at 1585cm$^{-1}$ ($I_{SERS}$) and the corresponding unenhanced signal from the bulk powder ($I_R$):

$$EF = \frac{I_{SERS}/N_{SERS}}{I_R/N_R}$$

where $N_{SERS}$ and $N_R$ are the estimated number of molecules contributing to SERS and Raman signals, respectively (Supp.Info.). From a spot size of 0.4μm and assuming that $N_{SERS}$=200 molecules are confined in each hotspot, we estimate the measured EF to be ~10$^8$ for this excitation wavelength. We compare this to predictions from numerical simulations of this geometry which suggest EF = $\left|E_p\right|^2|E_{SERS}|^2$=10$^6$- 10$^7$. By fitting Lorentzian lines to each vibrational peak and subtracting the SERS backgrounds, the spectral evolution $I_{SERS}(\nu, \lambda_p)$ of the three main TPT peaks is extracted. Normalising these to the incident laser power (we separately confirm all signals are linear in pump power) these are plotted as a function of the excitation wavelength and directly compared to the dark field spectrum of the same nanoparticle (Fig.2b).

**FDTD simulation**
The electromagnetic response of the nanoparticle on mirror geometry (NPOM) has been simulated by 3-dimensional finite difference time domain (FDTD) calculations using Lumerical FDTD Solutions v8.9. The structure has been modelled as a gold sphere of 60nm diameter on top of a 200nm thick gold layer, with a 1nm thick dielectric sheet in between. For the gold, we referred to the dielectric constants reported in Johnson and Christy.(35) The gold nanoparticle was illuminated with *p*-polarized plane waves from an angle of incidence of $\theta_i$ = 55°. The scattered light was then collected within a cone of half-angle $\theta_c$ = 53°, based on the numerical aperture of the objective.

## SUPPORTING INFORMATION

Laser power effects on the optical response of single nanoparticles, terphenylthiol powder absorption measurements, biphenyl-4-thiols tuneable SERS measurements on different size nanoparticles, FDTD simulations for a dimer geometry, transformation optics technique, scanning electron microscopy correlation.

## ACKNOWLEDGEMENT

We acknowledge financial support from EPSRC grants EP/G060649/1, EP/L027151/1, EP/G037221/1, EPSRC NanoDTC, and ERC grant LINASS 320503. J.A. acknowledges support from project FIS2013-41184-P from Spanish MINECO and project NANOGUNE'14 from the Dept. of Industry of the Basque Country. F.B. acknowledges support from the Winton Programme for the Physics of Sustainability. R.C. acknowledges financial support from St. John's College, Cambridge for Dr. Manmohan Singh Scholarship. P.A. acknowledges funding from the Helmholtz Association for the Young Investigator group VH-NG-928 within the Initiative and Networking fund. We thank Laurynas Pukenas and Steve Evans (University of Leeds, UK) for support with the ellipsometry measurements.